\documentclass[%
 reprint,
 amsmath,amssymb,
 aps,
]{revtex4-1}

\usepackage{graphicx} % Include figure files
\usepackage{dcolumn}  % Align table columns on decimal point
\usepackage{bm}       % bold math
\usepackage{verbatim} %add comments
\usepackage{gensymb}

\begin{document}

\preprint{APS/123-QED}

\title{Particle assembly with synchronized acoustical tweezers}% Force line breaks with \\

\author{Zhixiong Gong}
 %\altaffiliation[Also at ]{Physics Department, XYZ University.}
\author{Michael Baudoin}%
 \email{michael.baudoin@univ‑lille.fr}
\affiliation{%
 Univ. Lille, CNRS, Centrale Lille, ISEN, Univ. Polytechniques Hauts-de-France, UMR 8520-IEMN, International laboratory LIA/LICS, F-59000 Lille, France %\textbackslash\textbackslash
}%
\date{\today}% It is always \today, today,
\begin{abstract}
The contactless selective manipulation of individual objects at the microscale is powerfully enabled by acoustical tweezers based on acoustical vortices [Baudoin et al., Sci. Adv., 5:eaav1967 (2019)]. Nevertheless, the ability to assemble multiple objects with these tweezers has not yet been demonstrated yet and is critical for many applications, such as tissue engineering or microrobotics. To achieve this goal, it is necessary to overcome a major difficulty: the ring of high intensity ensuring particles trapping at the core of the vortex beam is repulsive for particles located outside the trap. This prevents the assembly of multiple objects. In this paper, we show (in the Rayleigh limit and in 2D) that this problem can be overcome by trapping the target objects at the core of two synchronized vortices. Indeed, in this case, the destructive interference between neighboring vortices enables to create an attractive path between the captured objects. The present work may pioneer particles precise assembly and patterning with multi-tweezers.
%\begin{description}
%\end{description}
\end{abstract}

\pacs{Valid PACS appear here}% PACS, the Physics and Astronomy
%\keywords{Suggested keywords}%Use showkeys class option if keyword display desired
\maketitle

%Part I: Introduction
\section{\label{sec:level1}Introduction}

Acoustical vortices \cite{hefner1999acoustical,marston2008scattering} have been attracting more and more attention for particles selective manipulation \cite{jap_baresch_2013,courtney2014independent,marzo2015holographic,thomas2017acoustical,riaud2017selective,marzo2019holographic,sa_baudoin_2019}. Unlike the ordinary zeroth-order Bessel beam \cite{durnin1987exact,durnin1987diffraction,marston2007scattering}, acoustical vortices are helical wave spinning around a phase dislocation in the azimuthal direction, wherein the amplitude cancels. This minimum of the beam intensity surrounded by a bright ring of high intensity can serve as an acoustical trap for particles more dense and/or more stiff than the surrounding fluid. Hence so-called cylindrical vortices (separate variable solution of Helmholtz equation in cylindrical coordinates \cite{marston2008scattering}) are well suited for 2D particles trapping (xy directions) \cite{courtney2014independent,riaud2017selective,fan2019trapping}. Nevertheless, since the intensity of these beams is invariant along the z axis, particles can only be pushed or pulled (in the so-called tractor beam configuration) \cite{marston2007negative,marston2009radiation,zhang2011geometrical,gong2017multipole,zhang2018general}, but not stably trapped in this direction. Instead, it was first proposed theoretically \cite{jap_baresch_2013} and then demonstrated experimentally \cite{baresch2016observation} by Baresh et al., that the use of spherical vortices enables to create 3D traps with a one-sided transducer. These wavefields can be obtained by adding an axial focalization to cylindrical vortices \cite{jap_baresch_2013}. Trapping of particles with such focalized vortices in air was also demonstrated by Marzo et al. \citep{marzo2015holographic}. Alternatively, Riaud et al. \cite{pre_riaud_2015} proposed to tame the degeneracy of a vortex beam between an anisotropic and an isotropic medium to create a 3D trap. More recently, it was shown both theoretically and experimentally \cite{phd_baresch_2014,marzo2018acoustic,baresch2018orbital}, that acoustical vortices can also be used to control the rotation of particles by using the pseudo-angular momentum carried by these helical structures. It is also noteworthy to underline that the calculations of the force and torque exerted by a vortex beam, initially calculated for spherical particles have been recently extended to nonspherical particles with the use of the T-matrix method \cite{gong2017multipole,gong2017t,gong2018thesis}.

Experimentally, the approaches for the synthesis of acoustical vortices can be divided into three categories: The first one, that we will refer as the \textit{active array method}, relies on arrays of transducers whose phase and/or amplitude can be tuned to synthesize a vortex in the surrounding fluid \cite{hefner1999acoustical,thomas2003pseudo,marchiano2008doing,courtney2014independent,riaud2015anisotropic,pre_riaud_2015,baresch2016observation,guild2016superresolution,ieee_riaud_2016, riaud2017selective,marzo2018acoustic,muelas2018generation}. The advantage of this method is that the vortex core (and thus the acoustical trap) can be moved electronically \cite{courtney2014independent,riaud2015anisotropic,ieee_riaud_2016,marzo2015holographic} and multiple traps can be synthesized simultaneously \cite{marzo2019holographic}. The disadvantage is that this technique relies on a complex array of transducers and a programmable electronics, which become expensive and complex to miniaturize as the frequency is increased to trap smaller particles (especially in liquids wherein the sound speed is 5 times larger that in air). The second one, the \textit{passive vortex-beam method} relies on passive devices, including spiral diffraction gratings \cite{jimenez2016formation,wang2016particle,apl_jimenez_2018}, metasurfaces or metastructure \cite{jiang2016convert,ye2016making,tang2016making}, which convert a plane or focused acoustic field into vortex beams. Finally, a third promising method, the \textit{active hologram method}, relies on the patterning of holographic electrodes at the surface of an active piezoelectric material. In this system the hologram of the vortex is directly engraved in the shape of the electrodes, which can be patterned with classic photolithography techniques. This last method enables to produce cheap, miniaturized, flat, transparent tweezers, which can be easily integrated in a classical microscopy environment to monitor displacement of small particles \cite{riaud2017selective,sa_baudoin_2019}.

It is important to note that the quest for the development of one-sided selective tweezers must be distinguished from the extensive work on acoustical traps based on standing  waves \cite{csr_lenshof_2010,loc_ding_2013,nm_ozcelik_2018}. Indeed, tweezers based on standing waves are versatile tools for particles sorting or collective manipulation of multiples objects, but the multiple nodes and antinodes prevent any selectivity. Indeed, since the wavefield is not localized in space (contrarily to vortex beams wherein the energy is focalized), it is not possible to move one particle independently of other neighboring particles. In addition, standing waves can only be obtained by placing some transducers (or reflectors) all around the area of interest, which prevents the design of one-sided 3D tweezers.

 While acoustical tweezers based on acoustical vortices have undergone tremendous progress toward miniaturization and improved selectivity on one side \cite{sa_baudoin_2019} and multiple objects simultaneous manipulation  on the other side \cite{marzo2019holographic}, one key operation is still missing for many applications, such as tissue engineering or micro-robotics: the ability to assemble multiple objects. At first sight this operation seems incompatible with the vortex beam structure. Indeed, the particles are trapped at the center of a high intensity ring which ensures the inner particle trapping, but which is repulsive for particles located outside the trap. Thus two particles located respectively inside and outside of the trap  cannot be assembled. In this paper, we demonstrate theoretically that this problem can be overcome by trapping the objects to be assembled at the center of two-synchronized vortex beams. Indeed, the destructive interference between these beams (when they are brought close to each other) creates an attractive path between the trapped particles. This demonstration is conducted in the limit of cylindrical vortices (transverse assembly) and in the long-wavelength approximation (i.e. for particles much smaller than the wavelength). With this calculation, we are also able to estimate the critical velocities for particle assembly depending on the properties of the trapped particles and surrounding fluid as well as the beam intensity.

%Part II: Theory and Mathematics (Using Mathpix for LaTeX codes of equations)
\section{\label{sec:level2B}Interaction between cylindrical vortices: intensity and phase}
%II.A \subsection{\label{sec:level2A}Interaction for different phase shift}
%  * means a total line for the Fig; Accept .EPS,.PNG,.JPG,.PDF
\begin{figure} [t]
\includegraphics[width=8.6cm]{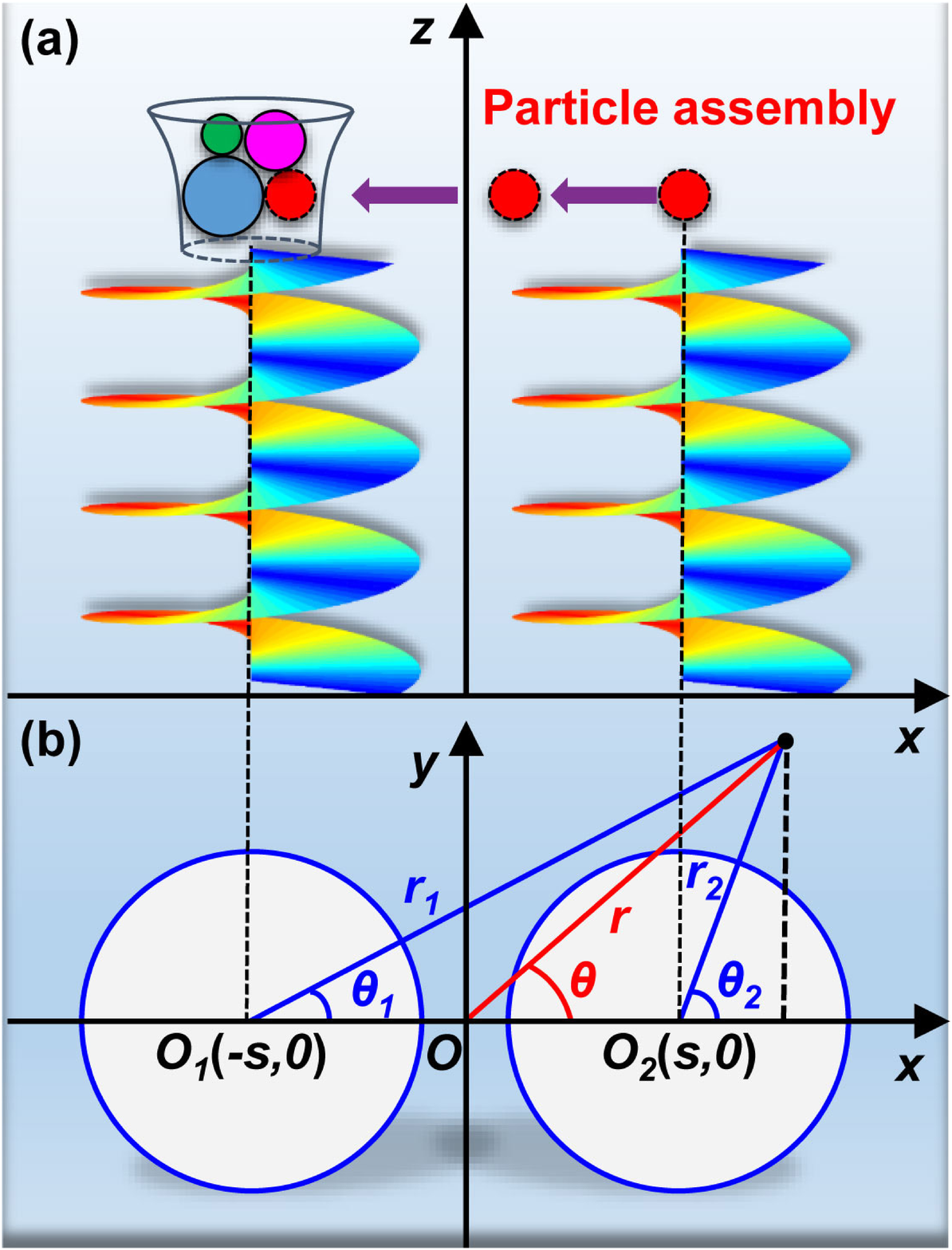}
\caption{Schematic of two cylindrical Bessel beams. (a) Particle assembly with two vortices. The particles can have different properties and size within the Rayleigh regime; (b) Geometrical relationship of radii and azimuthal angles in local and global coordinates systems.}
\label{Fig1:schematic}
\end{figure}
In this section we will consider the interference in a transverse plane of fixed altitude ($z=0$) between two cylindrical Bessel beams whose respective centers are located in $O_{1}(-s,0)$  and  $O_{2}(s,0)$). Hence, the distance between the original cores of these two vortices is $2s$ along the $x$ axis as shown in Fig.\ref{Fig1:schematic}. In this configuration, the pressure field $p_j$ produced by the $j$th vortex ($j=1$: left vortex; $j=2$: right vortex) is given by the equation:
\begin{equation}
p_{j}=A_{j} J_{m}\left(k_\perp r_{j}\right) e^{i m_{j} \theta_{j}} e^{i \beta_{j}},
\label{Eq1 Pressure}
\end{equation}
with $r_j$  the radial distance with respect to the origin of the $j$th vortex beam, $A_j$ the  beam amplitude, $m_{j}$ its topological charge, $\theta_j$  and $\beta_j$  the azimuthal and original phase angles, respectively and finally $k_\perp = k \sin(\gamma)$ the transverse wavenumber, with $k =\omega/c$ the wavenumber, $c$ the sound speed in the fluid, $\omega$ the angular frequency, and $\gamma$ the so-called cone angle. We can note that a cylindrical Bessel beam is entirely defined by its topological charge $m$ and the cone angle $\gamma$. For the sake of simplicity, we will consider vortices of topological order $m=1$ (the order mostly considered to design acoustical tweezers since it provides the stiffest trap) and $\gamma = 90 \degree$ (leading to $k_\perp = k)$. Present results can be easily extended to any values of $m$ and $\gamma$ by following the same procedure as the one expressed below.

\begin{figure*} [t]
\includegraphics[width=17.2cm]{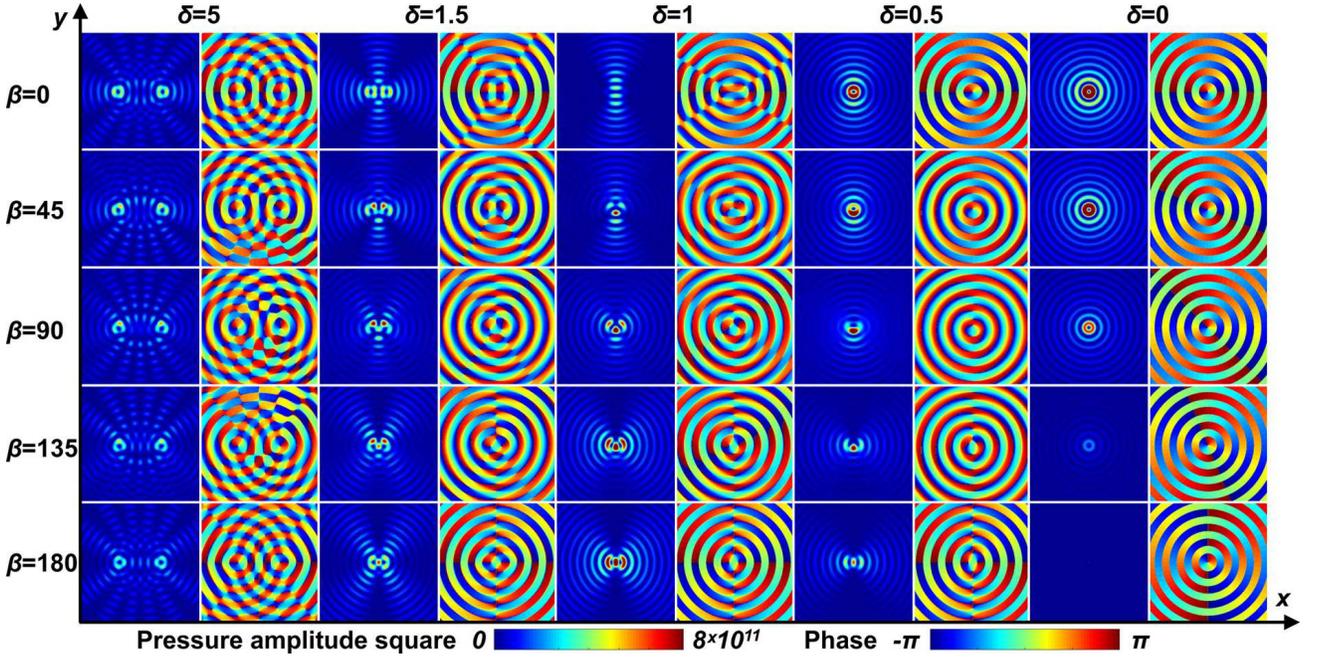}
\caption{Square of the pressure amplitude $|p|^{2}$ and phase distribution for two interfering cylindrical Bessel beams with different offset ratios $\delta=s/l_{0}$ (representing the distance between the cores of the two vortices)  and phase shifts ($\beta$ in degrees). The computational domain is $x \in [-4\lambda,4\lambda]$ and $y \in [-4\lambda,4\lambda]$ with $\lambda$ the wavelength. For each $\delta$, the left and right columns are amplitudes and phases of the interfering vortices, respectively. This picture shows that the most optimal case for particle assembly is the in-phase case ($\beta$=0). Movie M1 shows the continuous evolution of the phase and amplitude as a function of $\delta$ in this case.}
\label{Fig2:Amplitude and phase}
\end{figure*}

The total pressure field created by the interference of these two vortices is simply the sum $p=p_1+p_2$. For convenience, we will introduce the phase shift between these two vortices $\beta=\beta_2-\beta_1$ (set $\beta_1$=0, $\beta_2=\beta$ in previous equation) and introduce the local $(r_j, \theta_j)$ and global coordinates $(r,\theta)$ linked by the equation:
\begin{equation}
r_{1,2} e^{i \theta_{1,2}}=r e^{i \theta} \pm s.
\label{Eq2 geometrical relationship}
\end{equation}
The local coordinates are centered on each vortex central axis $j$ and the global coordinated are centered in between the axis of the two vortices (see Fig.\ref{Fig1:schematic}(b)). By substituting Eq.(\ref{Eq2 geometrical relationship}) into (\ref{Eq1 Pressure}), the total complex pressure reads:
\begin{equation}
\begin{aligned}
p=\left[A_{1}\frac{J_{1}\left(k r_{1}\right)}{r_{1}}+A_{2} \frac{J_{1}\left(k r_{2}\right)}{r_{2} }e^{i\beta}\right] r e^{i\theta}  \\
+\left[A_{1} \frac{J_{1}\left(k_{1}\right)}{r_{1}}-A_{2}\frac{J_{1}\left(k r_{2}\right)}{r_{2}}e^{i\beta}\right]s
\label{Eq3 Total pressure}
\end{aligned}
\end{equation}
It should be noted that the phase term of the sound velocity is no longer the same as the one of the total complex pressure for the synthetic field. For convenience, we will quantify the beam intensity with the value of $|p|^{2}$ and the phase of the total field is simply the argument of the complex pressure $arg(p)$. Fig.\ref{Fig2:Amplitude and phase} shows the square of the pressure amplitude $|p|^{2}$ and phase distributions of the superposition of two cylindrical Bessel beams as a function of the dimensionless offset ratio $\delta=s/l_0$ and the original phase shift $\beta$, with $s$ the distance between the beam axis $O_{1,2}$ and the origin $O$ (see Fig.\ref{Fig1:schematic}) and $l_0$ the distance between the maximum pressure amplitude on the first ring and the origin of a single vortex beam (corresponding to the position of the first maximum of the cylindrical Bessel function). The acoustic frequency is $f=1$ MHz and the beam amplitudes are $A_{1,2}=A=10^{6}$ Pa throughout the paper unless  mentioned otherwise. We can notice that this beam amplitude $A$ leads to a maximum value of the pressure on the first ring  $p_{max} = A \times J_1^{max} \approx 0.6 \times 10^{6}Pa$. The computational domain is $x \in [-4\lambda,4\lambda]$ and $y \in [-4\lambda,4\lambda]$, where $\lambda$=301 $\mu$m in water at the frequency considered in this paper with the acoustic parameter listed in Table \ref{Table1: Acosutic parameters}.

The results presented in Fig.\ref{Fig2:Amplitude and phase} show that: (i) The vortices cores (and hence tweezers traps) are located on the $x$ axis in the in-phase ($\beta=0 \degree$) and out of phase ($\beta=180 \degree$) cases, while they deviate from the individual core-core line for other phase shifts, such as $\beta=45 \degree$, $90 \degree$ or $135\degree$. Even in the in-phase case, the positions of the composite vortices cores (resulting from the interference between the two vortices) do not correspond to the positions of the axes of each individual vortex synthesized separately (as we will see in section \ref{sec:level2A}). These results are consistent with the work of Maleev \& Groover \cite{maleev2003composite}, who studied the phase singularities resulting from the interaction of two optical vortices. (ii) When the two vortices axes coincide ($\delta = 0$), the interference pattern evolves entirely from the constructive interference in the in-phase case ($\beta = 0\degree$), wherein the ring amplitudes sum-up, and the completely destructive case when the two vortices are in opposition of phase ($\beta = 180 \degree$), leading to a complete cancelling of the pressure amplitude. (iii) The in-phase case ($\beta = 0 \degree$) is the most suited configuration for particle assembly since it produces the largest and most isotropic rings surrounding the two vortices cores. (iv) Most importantly, the lateral destructive interference between two in-phase vortex beams when $\delta \rightarrow 1$, creates a path between the repulsive rings which might enable the assembly of two particles (see Movie M1 for a continuous description of the evolution of the amplitude and phase as a function of $\delta$). Nevertheless, since the situation $\delta \rightarrow 1$ also leads to the weakest pressure gradient in the $x$ direction, further investigation is still necessary to determine whether this gradient is sufficient to maintain a trap along this direction and determine the limited speed at which two particles could be assembled in this configuration.

%Part III: Results
%III.A
\section{\label{sec:level3A }Particles trapping with synchronized vortices}
For this purpose, we will examine the forces exerted on particles trapped at the core of two synchronized  acoustical vortices ($\beta = 0 \degree$) based on Gor'kov \cite{Gorkov1962on} expression of the radiation force in the long wavelength regime. Indeed, Gor'kov demonstrated that this force can be expressed, for a standing wave, as the gradient of a potential, now referred to as Gork'ov potential. In this section, we will compute this potential and the resulting lateral forces (in the $x$ direction) exerted on two particles initially trapped at the center of two cylindrical acoustical vortices. We will then compare this force to the Stokes drag in order to determine the critical velocity $v_{cr}$ at which two particles can be assembled by moving the two tweezers laterally. Indeed, if the Stokes drag exceeds the radiation force, the particles can be expelled from the trap. Finally, we will perform a parametric study of this critical velocity for different typical elastic materials and fluids.
\begin{table}[b]
\caption{Acoustic parameters for typical elastic materials and various fluids. The properties of polystyrene (PS), Pyrex, a representative biological cell (Cell), water and glycerol are taken from Ref. \cite{settnes2012forces}, the ones for ethanol from The Engineering ToolBox. The viscosities of blood plasma (BP) and ethanol are from Ref. \cite{bui2018calibration}, density from \cite{blooddensity} and sound speed from Ref. \cite{bloodspeed}.}
\begin{tabular}{c c c c c}
\hline
\hline
\begin{tabular}[c]{@{}c@{}}Material\end{tabular}                            & 
\begin{tabular}[c]{@{}c@{}}Density\\ $\rho_0$ ($kg/m^3$)\end{tabular}         & 
\begin{tabular}[c]{@{}c@{}}Compress.\\ $\kappa$ (1/TPa)\end{tabular}       & 
\begin{tabular}[c]{@{}c@{}}Longitud. speed\\of sound $c$ (m/s)\end{tabular} & 
\begin{tabular}[c]{@{}c@{}}Viscosity\\ $\eta$ (mPa s)\end{tabular}          \\
\hline
PS       & 1050   & 172   & 2350   & ---          \\
Pyrex    & 2230   & 27.8  & 5674   & ---          \\
Cell     & 1100   & 400   & 1500   & ---          \\
Water    & 998.2  & 456   & 1482   & 1.002        \\
BP       & 1025   & 407   & 1549   & 1.43         \\
Ethanol  & 785    & 973   & 1144   & 1.10         \\
Glycerol & 1261   & 219   & 1904   & 1412         \\
\hline 
\hline
\end{tabular}
\label{Table1: Acosutic parameters}
\end{table}

\begin{figure*} [htb]
\includegraphics[width=17.2cm]{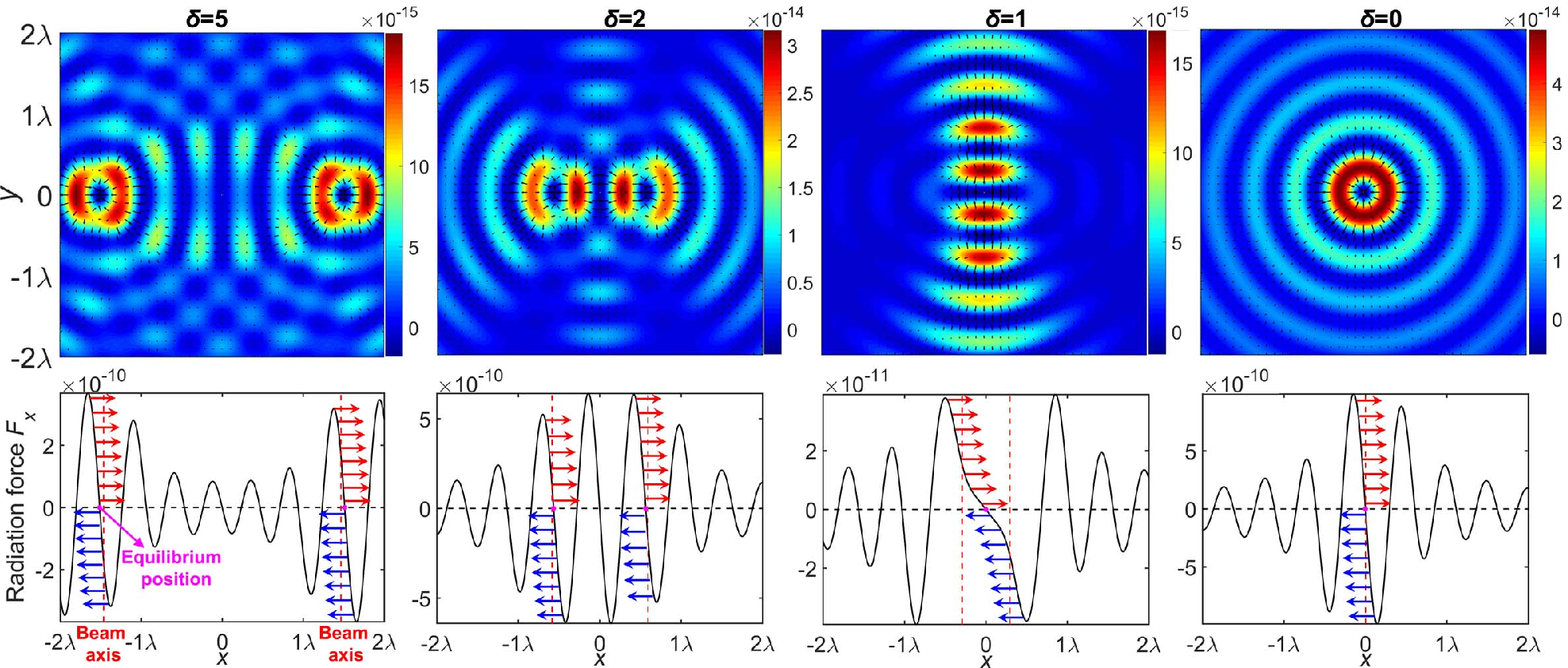}
\caption{Upper rows: Gor'kov potential (color) and radiation forces (arrows) resulting from the interaction of two interacting acoustical vortices (of amplitudes $A = 10^6$ Pa and driving frequency $f = 1$ MHz) with PS spheres of $5 \; \mu m$ for different offset ratios $\delta = s / l_o$ (dimensionless distance between the two vortices). Lower row: Lateral forces $F_x$ applied on the particles as a function of $x$. The static equilibrium positions (traps) correspond to the null force values in regions of negative force gradient $(\nabla F_x < 0)$. The individual vortex cores (in absence of interference) are represented with a red dashed line. The movie M2 in the supplementary material shows the continuous evolution of Gor'kov potential and radiation forces from $\delta=5$ to $\delta=0$.}
\label{Fig3:Gorkov potential and forces}
\end{figure*}

\subsection{\label{sec:level2A}Gor'kov potential and force}
For the estimation of the force, we will first consider some polystyrene (PS) particles of small radius $a=5 \; \mu m$ compared to the wavelength $\lambda = 300 \; \mu m$, in the range of validity of Gork'ov theory ($ka \approx 1/10 \ll 1$). These particles are chosen to demonstrate that particle assembly is possible with this method even for low acoustical contrast with the surrounding fluid (see acoustic parameters listed in Table \ref{Table1: Acosutic parameters}). Following Gor'kov \cite{Gorkov1962on}, the radiation force $\mathbf{F}$ can be expressed as the negative gradient of the potential $U$,   $$\mathbf{F}=-\mathbf{\nabla} U,$$ with:
\begin{equation}
U=2 \pi a^{3} \rho_0\left(\frac{\left\langle p^{2}\right\rangle}{3 \rho_{0}^{2} c^{2}_0} f_{1}-\frac{\left\langle \mathbf{v}^{2} \right\rangle}{2} f_{2}\right),
\label{Eq4 Gor'kov potential}
\end{equation}
and $f_{1}=1-\tilde{\kappa}$ (with the compressibility ratio $\tilde{\kappa}=\kappa_p/\kappa_0$) and $f_{2}=2\left(\tilde{\rho}-1\right) /\left(2 \tilde{\rho}+1\right)$ (with density ratio $\tilde{\rho}=\rho_p/\rho_0$) are the respective contributions of the monopolar and dipolar vibrations of the particle \cite{bruus2012acoustofluidics}. The compressibility of an elastic particle is $\kappa_p=1/K_p$ with $K_p=\rho_{p}\left(c_{l}^{2}-4 / 3 c_{t}^{2}\right)$ the bulk modulus, while the compressibility of the fluid is $\kappa_0=1/K_0$ with $K_0=\rho_{0} c_{0}^{2}$ the bulk modulus of the fluid. In these equations, $\rho_0$ and $\rho_p$ represent the fluid and particles density, $c_0$ the acoustic velocity in the fluid and $c_l$, $c_t$ the longitudinal and transverse velocities of elastic particles, and finally $p$ and $\mathbf{v}$ the pressure and velocity associated with the total incident acoustic field in the fluid. Note that the expression of $f_1$ in Gor'kov's original paper \cite{Gorkov1962on} based solely on the longitudinal velocity inside the particle only applies for liquid spheres. In the Fourier space, the velocity vector is related to the complex pressure, as $\mathbf{v}=-i /(\rho \omega) \nabla p$ with $\nabla= \frac{\partial}{\partial r} \;  \mathbf{e_r}+ \frac{1}{r} \frac{\partial}{\partial \theta} \;  \mathbf{e_\theta}$ the gradient operator in cylindrical coordinates, leading for each vortex to:
% & means align;\left,\right\\可以在公式中间换行
%*使latex不自动显示序号，如果想让latex自动标上序号，则把*去掉
\begin{equation}
\begin{aligned}
\begin{array}{r}
{\mathbf{v}_{j}=-i \frac{1}{\rho \omega} \nabla p_{j}=-i A_{j} e^{i \beta_{j}} \frac{1}{\rho \omega} \left. \Big\{ \frac{\partial}{\partial r}\left[J_{1}\left(k r_{j}\right) \frac{r e^{i \theta} \pm s} {r_{j}}\right] \mathbf{e_r} \right.} \\[2mm] %add the line space 2 mm
+{\frac{\partial}{r\partial \theta}\left[J_{1}\left(k r_{j}\right) \frac{r e^{i \theta} \pm s}{r_{j}}\right] \mathbf{e_\theta}} \Big\}.
\end{array} 
\end{aligned}
\label{Eq5 Velocity}
\end{equation}
In this expression, the sign before $s$ is $+$ for $j$=1 and $-$ for $j$=2. As previously, if we set $\beta_{1}=0$, we have $\beta_{2}=\beta$. To compute the derivative of the expressions in Eq.(\ref{Eq5 Velocity}) with respect to $r$ and $\theta$, we use the trigonometric relations (see Fig.\ref{Fig1:schematic}):  $r_{1}^{2}=r^{2}+s^{2}+2 r  s  \cos \theta$ and $r_{2}^{2}=r^{2}+s^{2}-2 r  s  \cos \theta$. Substituting Eqs.(\ref{Eq3 Total pressure}) and (\ref{Eq5 Velocity}) into the expression of Gor'kov potential in Eq. (\ref{Eq4 Gor'kov potential}), leads to the final expression of the force given in Appendix A.

The 2D Gor'kov potential and its corresponding negative gradient for two synchronized vortices are depicted in the first row of Fig.\ref{Fig3:Gorkov potential and forces} for offset ratios $\delta$=5, 2, 1 and 0. These figures show that particles can be trapped separately in the two vortices, when the offset is relatively large and can also be assembled in the center of the two coaxial beams when $\delta$=0. To further demonstrate the  feasibility of the two synchronized vortices structure for particle assembly, the lateral Gor'kov forces in the $x$ direction are calculated and presented in the bottom row of Fig.\ref{Fig3:Gorkov potential and forces}. The particles get trapped when the static equilibrium positions (zero force) are located in the region of negative forces gradient, with the restoring forces pushing them back to the equilibrium positions. This can be observed on Fig.\ref{Fig3:Gorkov potential and forces}, where the red arrows describe positive forces while the blue ones the negative forces, which both push the particles (magenta solid spheres) back to the trapping well. We can note that these static equilibrium positions (marked as magenta solid circles) differ from the position of the individual vortex axes (marked as red dashed lines) when the two vortex beams are not coaxial, due to the interference between the two vortices.  The Gor'kov potential and dynamic motion of the equilibrium positions for offsets ranging from $\delta=5$ to $\delta=0$ are further shown in the supplementary movie M2, which demonstrates the ability of the synchronized vortices to assemble particles.

%III.B
\subsection{\label{level3B}Gor'kov and critical Stokes' drag forces}
In the previous section, we calculated the quasi-static equilibrium positions of the particles (when the velocities of the tweezers traps tends toward $0$). In practice, it is of course necessary to move particles with a finite velocity, which leads to the existence of a drag force applied by the fluid on the particles. This drag force can expel the particle from the trap if the radiation force is not sufficient to counteract it. Hence, it is essential to determine what is the speed limit at which two particles can be assembled. To compute this critical velocity, we will assume (i) that the particles are moved by the tweezers at a constant velocity $v_{cr}$ and (ii) that the flow around the particles is in the low Reynolds regime ($Re \ll 1$). Considering the small size of the particles ($a = 5 \; \mu m$), this assumption holds for particles velocities: $v_{cr} \ll \eta / \rho a = 0.2 \; m s^{-1}$ in water, with $\eta$ the fluid dynamic viscosity. This value is several orders of magnitude larger than the typical velocities used for microparticles manipulation and thus, the low Reynolds hypothesis is consistent. Under this circumstance and assuming spherical particles, the magnitude of the drag force is simply the Stokes drag: $F_{d}=6 \pi \eta a v_{cr}$. 

\begin{figure} [htb]
\includegraphics[width=8.6cm]{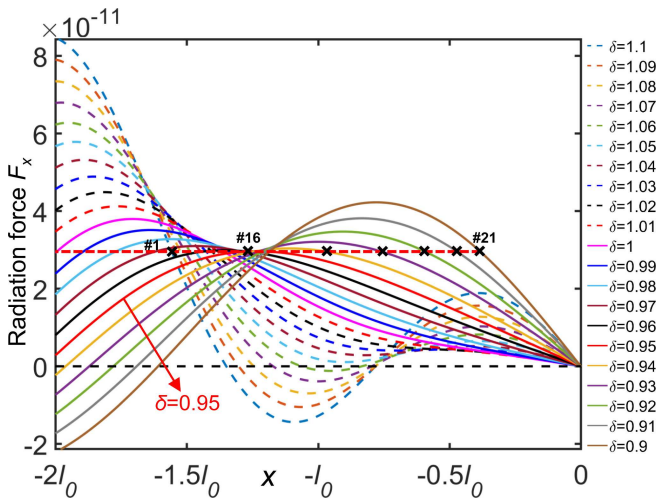}
\caption{Lateral radiation force $F_x$ applied on PS spheres immersed in water by two interfering vortex beams with offset ratios ranging from $\delta=1.1$ to $\delta=0.9$. The graph only shows the force for particles located in the half plane $x<0$ on the x axis. Results for particles located in the half plane $x>0$ can be inferred from these data by simple symmetry. The red dashed line describes the critical radiation force at $\delta=0.95$ which is balanced by the Stokes drag force to determine the critical velocity $v_{cr}$ of the moving tweezers.}
\label{Fig4:critical velocity}
\end{figure}

This drag force must be compared to the trapping force calculated in the previous section. As can be observed in Fig.\ref{Fig3:Gorkov potential and forces}, the lateral trapping force in the $x$ direction becomes minimal when $\delta \approx 1$. To determine more precisely this critical value of the offset, we computed the radiation force along x axis $F_{x}$ versus the $x$ coordinates in the half plane $x\in [-2l_{0},0]$, with the offset ratios ranging from $\delta=1.1$ to $\delta=0.9$ with increment $\Delta\delta=0.01$, as shown in Fig. \ref{Fig4:critical velocity} (the values of $F_x$ in the other half plane can be obtained by symmetry). Based on our theoretical computations, it is found that the minimum lateral trapping force peak is obtained for $\delta$=0.95. The maximum critical velocity at which two particles can be assembled (assuming that the particles move at a constant speed) is thus obtained when the drag force is equal to the value of the radiation force along x axis $F_x^{\mbox{\tiny{peak}}}$ on the peak situated on the left of the static equilibrium position (represented by the red dashed line on Fig.\ref{Fig4:critical velocity}): $F_{d} = F_x^{\mbox{\tiny{peak}}} (\delta = 0.95)$. Indeed, this is this peak which prevent the particle to escape from the trap when the tweezers are approached. This leads to the critical velocity:
\begin{eqnarray}
v_{cr} = F_x^{\mbox{\tiny{peak}}} (\delta = 0.95) / 6 \pi \eta \label{vcr}.
\end{eqnarray}
Below this speed, the particle can be trapped for all offset ratios, since the trapping force is always sufficient to counteract the Stokes drag. On Fig.\ref{Fig4:critical velocity}, the successive dynamic equilibrium positions of one of the trapped particles at this critical speed (numbered from $\#1$ to $\#21$) are represented by black crosses, when the offset is reduced from $\delta=1.1$ to $\delta=0.9$. As observed, the particle can be successively moved and then brought to a single centered trap, when the two vortex beams approach one another.

%III.C
\subsection{\label{level3C:input power}Critical velocity: Parametric study}
\begin{figure}[htbp]
\includegraphics[width=8.6cm]{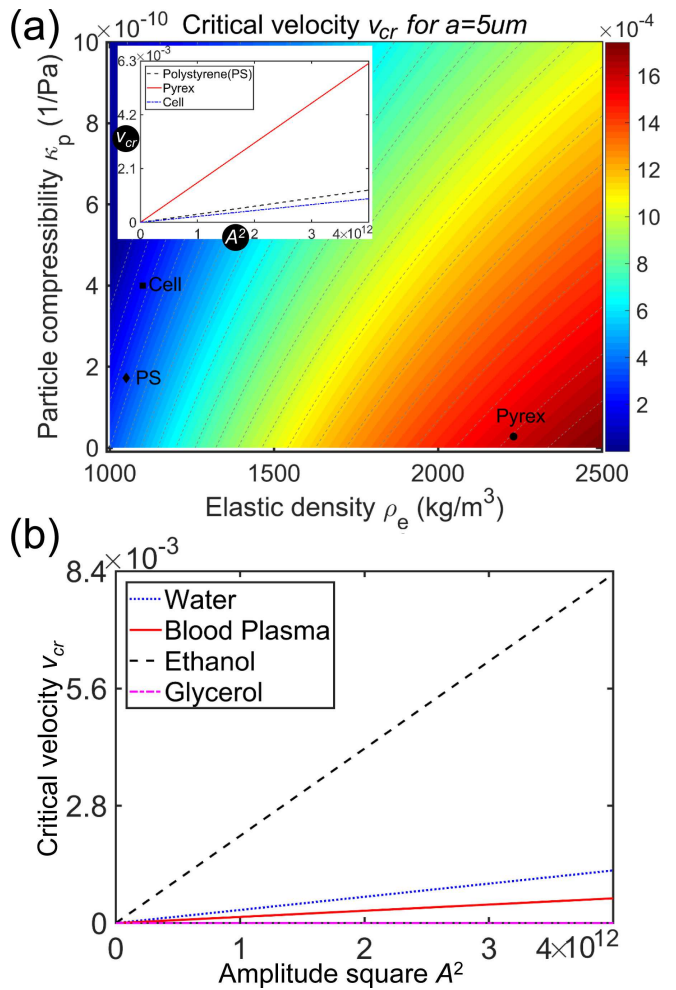}
\caption{(a) Critical velocity for the assembly of  5 $\mu$m elastic sphere embedded in water as a function of the particle density and compressibility at a fixed pressure amplitude $A=10^{6}$ and driving frequency $f = 1$ MHz. Insert: Critical velocity for four specific materials as a function of the square of the acoustic wave pressure amplitude $A^{2}$. (b) Critical velocity for a 5 $\mu$m polystyrene (PS) sphere embedded in different liquids as a function of the square of the acoustic wave pressure amplitude $A^{2}$.}
\label{Fig5:input power and critical velocity}
\end{figure}

In the previous section \ref{level3B}, we determined the expression of the critical velocity for particle assembly with two synchronized vortices for PS spheres. We will now extend these results to different particles, fluids and input power, with the acoustic parameters listed in TABLE \ref{Table1: Acosutic parameters}. In particular we will determine from Eq. (\ref{vcr}) the critical velocity for the following particles: PS, pyrex, a representative biological cell (Cell) and surrounding fluids: water, ethanol, blood plasma (BP) and glycerol. The evolution of the critical velocity $v_{cr}$ as a function of the input power $A^{2}$ for different particles in water is represented on the insert of Fig.\ref{Fig5:input power and critical velocity}(a). The beam amplitude varies from $A=0$ to $A=2\times10^{6}$. Obviously, the critical velocity is proportional to $A^{2}$. Indeed, Gor'kov potential in Eq.(\ref{Eq4 Gor'kov potential}) depends on the terms $\left\langle p^{2}\right\rangle$ and $\left\langle \mathbf{v}^{2}\right\rangle$, which are both proportional to the square of the beam amplitude $A$ (see Eqs. (\ref{Eq3 Total pressure})) and (\ref{Eq5 Velocity}). Then since the critical velocity can be expressed as $v_{cr}=-\partial{U}/\partial{x}(\delta = 0.95) /(6 \pi \eta a)$, it is also proportional to $A^{2}$. The values of the critical velocities for the assembly of small PS and pyrex particles and typical cells in water with a driving pressure amplitude $A$ of $2 \; MPa$ are compatible with experiments in microchannels. To broaden this analysis to any type of particles, we plotted a 2D graph (Fig.\ref{Fig5:input power and critical velocity}(a)), featuring the values of the critical velocity in water as a function of the particle compressibility and density at the fixed frequency $f$=1 MHz and amplitude $A = 1 \; MPa$. These results show, as expected, that cells are one of the most difficult objects to address, owing to their low acoustic contrast. Nevertheless, they can still be moved and assembled with reasonable speeds of several hundreds microns per second at acoustic levels that remain harmless for living objects (far below the cavitation threshold and power levels used e.g. for echography). Finally, we computed the critical velocity for PS particles embedded in different common fluids (Fig.\ref{Fig5:input power and critical velocity}(b)) as a function of the driving pressure. This time, three properties of the fluid influence the critical velocity: the acoustic wave speed, the fluid density influencing the radiation force, and the viscosity of the fluid influencing the Stokes drag.

%Part IV: Conclusion
\section{\label{level4:Conclusion}Discussion and conclusion}
In this paper, we demonstrated theoretically that two synchronized cylindrical vortex beams are suitable to move and assemble particles with sizes much smaller than the wavelength in 2D. Indeed, the interference between two neighboring vortex beams creates an attractive path between the two initial rings used for particles trapping. This enables merging the particles without the need to cross the rings potential barrier. The maximum critical velocities for particle assembly are determined based on the balance between the radiation and drag forces. Indeed, the former maintains the particle in the trap, while the latter tends to push it out of the trap. The estimated speeds are compatible with microparticles manipulation in microchannels, even for particles with low contrast such as cells, at acoustic levels compatible with the manipulation of living objects. This work constitutes a first step toward particle assembly with acoustical tweezers. Potential continuation of this work includes the extension to 3D particle assembly, in particular beyond the long wavelength regime and the consideration of particles rotation.

%Part V
\section*{\label{sec:Acknowledgmnts}Acknowledgments}
We acknowledge the support of the programs ERC Generator and Prematuration funded by ISITE Universit\'{e} Lille Nord-Europe.

%Part VI Appendix
\section*{\label{sec:Appendix}Appendix: Expressions of velocity and radiation force}
\setcounter{equation}{0}
\renewcommand\theequation{A.\arabic{equation}}

The velocity expressions of the individual vortex (set $\beta_1$=0, $\beta_2=\beta$ throughout the paper) are:   
\begin{equation}
\begin{aligned}
\begin{array}{r}
{\mathbf{v}_{1}=-i \frac{1}{\rho \omega} A_{1}\left\{C_{1} \mathbf{e_r}+B_{1} \mathbf{e_\theta}\right\}}.
\end{array} 
\end{aligned}
\label{v1 Velocity}
\end{equation}
and 
\begin{equation}
\begin{aligned}
\begin{array}{r}
{\mathbf{v}_{2}=-i \frac{1}{\rho \omega} A_{2} e^{i \beta}\left\{C_{2} \mathbf{e_r}+B_{2} \mathbf{e_\theta}\right\}}.
\end{array} 
\end{aligned}
\label{v2 Velocity}
\end{equation}
with the coefficients $C_1$, $B_1$, $C_2$ and $B_2$

\begin{equation}
\begin{aligned}
\begin{array}{r}
{C_1=\frac{\left[J_{1}^{\prime}\left(k r_{1}\right)  k r_{1}-J_{1}\left(k r_{1}\right)\right](r+s  \cos \theta)}{r_{1}^{3}} \left(r e^{i \theta}+s\right)+
\frac{J_{1}\left(k r_{1}\right)}{r_{1}} e^{i \theta}},
\end{array} 
\end{aligned}
\label{C1 Velocity}
\end{equation}

\begin{equation}
\begin{aligned}
\begin{array}{r}
{B_1=\frac{-\left[J_{1}^{\prime}\left(k r_1\right)  k r_{1}-J_{1}\left(k r_{1}\right)\right]s  r \sin \theta}{r_{1}^{3}} \left(e^{i \theta}+\frac{s}{r}\right)+
\frac{J_{1}\left(k r_{1}\right)}{r_{1}}  i  e^{i \theta}},
\end{array} 
\end{aligned}
\label{B1 Velocity}
\end{equation}

\begin{equation}
\begin{aligned}
\begin{array}{r}
{C_2=\frac{\left[J_{1}^{\prime}\left(k r_{2}\right)  k r_{2}-J_{1}\left(k r_{2}\right)\right](r-s  \cos \theta)}{r_{2}^{3}}\left(r e^{i \theta}-s\right)+
\frac{J_{1}\left(k r_{2}\right)}{r_{2}} e^{i \theta}},
\end{array} 
\end{aligned}
\label{C2 Velocity}
\end{equation}

\begin{equation}
\begin{aligned}
\begin{array}{r}
{B_2=\frac{\left[J_{1}^{\prime}\left(k r_{2}\right)  k r_{2}-J_{1}\left(k r_{2}\right)\right]s  r \sin \theta}{r_{2}^{3}} \left(e^{i \theta}-\frac{s}{r}\right)+
\frac{J_{1}\left(k r_{2}\right)}{r_{2}}  i e^{i \theta}}.
\end{array} 
\end{aligned}
\label{B2 Velocity}
\end{equation}

The total velocity vector is $\mathbf{v}=\mathbf{v}_{1}+\mathbf{v}_{2}=v_{r} \mathbf{e_r}+v_{\theta} \mathbf{e_\theta}$, with the time average 
\begin{equation}
\begin{aligned}
\begin{array}{r}
{\left\langle\mathbf{v}^{2}\right\rangle=\frac{1}{2} \operatorname{Re}\left\{v_{r} v_{r}^{*}+v_{\theta} v_{\theta}^{*}\right\}}.
\end{array} 
\end{aligned}
\label{A3 Velocity}
\end{equation}
where $v_{r}=-i \frac{1}{\rho \omega}*\left(A_{1} C_{1}+A_{2} e^{i \beta} C_{2}\right)$ and $v_{\theta}=-i \frac{1}{\rho \omega}*\left(A_{1} B_{1}+A_{2} e^{i \beta} B_{2}\right)$. The time average of pressure square could be expressed as $\left\langle p^{2}\right\rangle=\frac{1}{2} \operatorname{Re}\left\{p p^{*}\right\}$. The force can be simply obtained by substituting the above equations into Gorkov's potential in Eq.(\ref{Eq4 Gor'kov potential}).

%Refs: All the Refs need to be cited so it will be correct in the paper
\renewcommand\refname{Reference}
\bibliographystyle{unsrt}  %Set bibliography style unsrt:rank as the citation order
\bibliography{Refs}        %Produce the bibliography

\end{document}